# Electrocaloric effect and high energy storage efficiency in lead-free $Ba_{0.95}Ca_{0.05}Ti_{0.89}Sn_{0.11}O_3$ ceramic elaborated by sol-gel method

Youness Hadouch[1, 2, *], Said Ben Moumen[1], Hanane Mezzourh[1, 2], Daoud Mezzane[1, 2], M'barek Amjoud[1], Bouchra Asbani[2], Anna G. Razumnaya[2,4], Yaovi Gagou[2], Brigita Rožič[3], Zdravko Kutnjak[3], Mimoun El Marssi[2].

*1 Laboratory of Innovative Materials, Energy and Sustainable Development (IMED), Cadi-Ayyad University, Faculty of Sciences and Technology, BP 549, Marrakech, Morocco.*
*2 Laboratory of Physics of Condensed Matter (LPMC), University of Picardie Jules Verne, Scientific Pole, 33 rue Saint-Leu, 80039 Amiens Cedex 1, France.*
*3 Jozef Stefan Institute, Jamova Cesta 39, 1000 Ljubljana, Slovenia.*
*[4] Faculty of Physics, Southern Federal University, Rostov-on-Don, 344090, Russia.*

*Corresponding author:

E-mail: hadouch.younes@gmail.com; youness.hadouch@edu.uca.ma

ORCID: https://orcid.org/0000-0002-8087-9494

Tel: +212-6 49 97 06 74

**Abstract:**

Structural, dielectric, ferroelectric, energy storage properties, and electrocaloric effect were studied in lead-free ceramic $Ba_{0.95}Ca_{0.05}Ti_{0.89}Sn_{0.11}O_3$ (BCTSn) elaborated by sol-gel method. Phase purity structure was confirmed from X-ray data using Rietveld refinement analysis which revealed the coexistence of tetragonal (P4mm) and orthorhombic (Amm2) symmetries at room temperature. Phase transitions were detected by dielectric and differential scanning calorimetry results. Energy storage properties were determined from P-E hysteresis, and the electrocaloric properties were calculated indirectly via the Maxwell approach. The large value of electrocaloric temperature change of $\Delta T=0.807$ K obtained at a relatively small field of 30 kV cm$^{-1}$ and high energy storage efficiency can make BCTSn ceramic a promising candidate for environmentally friendly refrigeration and energy storage applications.

## 1. Introduction:

One of the most critical issue in the energy consumption is the rising power consumption for air conditioning. The growing demand for cooling and air-conditioning machines increases the world's power consumption. However, most of these devices rely on vapor-compression technology, which emits harmful gases as fluorine. For that reason, environmentalists and scientists are moving to develop new technologies for environmentally friendly refrigeration systems. The electrocaloric effect is an innovative alternative technology based on the change of temperature (ΔT) and entropy (ΔS) in polar materials under an external electric field [1].

Ferroelectric ceramics have gained increasing attention because of their uses in various applications such as data storage, piezoelectric energy harvesting, and cooling devices [2–4]. Despite their robust properties, Pb-based materials face notable global restrictions, hence why it is necessary to develop new environment-friendly materials. Barium titanate $BaTiO_3$ (BT) is a typical alternative system that can become an important candidate for this application [5]–[7]. Nevertheless, it shows a low dielectric constant, high Curie temperature ($Tc$ = 120 °C), and low energy storage properties [8] that calls for strategies to improve these characteristics[9]. Site engineering as solution consists to substitute A-sites($Ba^{2+}$) with $Ca^{2+}$, $Sr^{2+}$, $La^{2+}$… and/or B-sites($Ti^{4+}$)with $Sn^{4+}$, $Zr^{4+}$, $Hf^{4+}$ [10]. More than 10 years ago, et al. [11]studied the effect of doping $Ba^{2+}$ with $Ca^{2+}$ in $Ba_{1-x}Ca_xTiO_3$ ceramics and reported an improvement of $d_{33}$ from 180 up to 310 pC/N for 0.02<x<0.34. In this context, several studies highlighted the partial substitution of Ti in $BaTiO_3$ by $Sn^{4+}$, and its effect on the phase transition dielectric behavior and electrocaloric properties [7], [12], [13]. It is well known that pure BT shows three distinct phase transitions with increasing temperature, i.e., the rhombohedral (R3m) – orthorhombic (Amm2) phase transition at $T_{R-O}$ = -90 °C, the orthorhombic – tetragonal (P4mm) phase transition($T_{O-T}$ =0 °C), and the tetragonal-cubic phase transition ($T_C$= 120 °C)[14]. The addition of Sn can shift $T_{O-T}$ and $T_{R-O}$ to room temperature [12]. Meanwhile, doping of B sites with $Ca^{2+}$(up to 21%) can reduce these transitions temperatures with a slight change on the Curie point [15]. Accordingly, co-doping in both sites can lead to the formation of the morphotropic phase boundary (MPB) at ambient temperature resulting in the enhancement of dielectric, piezoelectric and electrocaloric properties [7], [16], [17]. Theoretically, in the MPB region, energy barriers become lower, making the polarization rotation and extension process easier [18]. Consequently, it is the right choice for promoting electrical properties to create two-phases or multiphase coexistence near room temperature. Wang et al. studied the electrocaloric effect (ECE) in the $Ba_{0.94}Ca_{0.06}Ti_{1-x}Sn_xO_3$ system with multiphase coexistence and achieved a large ECE of ΔT = 0.63 K under 20 kV $cm^{-1}$ [19]. Benefiting from the MBP in BCTSn, Zhu et al reported that $(Ba_{0.95}Ca_{0.05})(Ti_{1-x}Sn_x)O_3$ compositions displayed a high piezoelectric responses with $d_{33}$=670 pC/N at x=0.11 [20]. Nevertheless, there has been no investigations on the energy storage performance of the BCTSn materials. However, some recent studies reported a high energy storage density at the polymorphic phase boundaries. For example, Merselmiz et al. [21]

achieved improved recovered energy storage (Wrec) of 72.4 mJ/cm$^3$ with a high energy storage efficiency η = 85.07 % in the BTSn ceramic under 25kV/cm. In addition, energy performances in Ba$_{0.85}$Ca$_{0.15}$Zr$_{0.10}$Ti$_{0.90}$O$_3$ with MPB were frequently studied and reached a large Wrec of 414.1 mJ cm$^{-3}$, with electrocaloric responsivity (η=78.6%) and a significant coefficient of performance (COP = input power/output cooling power) of 6.29 under 55 kV cm$^{-1}$ [22]. On the other hand, the synthesis method is an important factor for designing ceramics with high compositional homogeneity, error-free stoichiometric ratio and lower calcination and/or sintering temperature due to the mixture of precursors at the molecular level. The solid-state reaction route used to prepare BT-based ferroelectric materials necessitates a high calcination temperature (1350°C) and sintering temperature range of 1400-1550°C [23], [24]. Wang et al. [25] carried out a comparative study in BCZT elaborated by sol-gel (BCZT-SG) and solid-state (BCZT-SS) methods and reported a very high energy density of 0.52J cm$^{-3}$ in BCZT-SG comparing with BCZT-SS (0.32 J cm$^{-3}$). Furthermore, many researchers have argued that grain size affects the physicochemical properties of ferroelectric materials [26]–[28].

In this paper, we have experimentally explored the temperature-dependence of the energy storage and electrocaloric properties in the Ba$_{0.95}$Ca$_{0.05}$Ti$_{0.89}$Sn$_{0.11}$O$_3$ ceramic (BCTSn) elaborated by sol-gel process and sintered at 1350°C for 7h. The electrocaloric effect was evaluated using the indirect experimental approach following Maxwell relation. The enhanced values of ECE, electrocaloric responsivity (η), COP and energy storage were compared with literature data, suggesting that our BCTSn ceramic could be a promising candidate for environmentally friendly refrigeration and high energy storage applications.

**2. Materials and methods:**

The XRD patterns of BCTSn ceramic were obtained by X-ray diffraction using the Panalytical X-Pert Pro with Cu-Kα radiation (λ = 1.54059 Å) at room temperature. The apparent density of ceramic samples was determined by Archimedes method while the relative density was estimated from crystal structure parameters. The surface morphology of ceramic was examined using a scanning electron microscope (SEM, VEGA 3-Tescan). The temperature dependence of dielectric properties was measured using an impedance analyzer (HP 4284A) at the temperature, and frequency ranges -50 to 200°C and 20 Hz-100 kHz, respectively. Differential scanning calorimetry (Perkin Elmer Jade DSC) was used to point out the different phase transitions. The polarization-electric field (P–E) hysteresis loops were performed by CPE1701,

PloyK, USA, with a high voltage power supply (Trek 609-6, USA) in a silicone oil bath at 20 Hz in the temperature range of 30-120 °C. The electrocaloric study was carried out by the indirect method using the recorded P–E hysteresis loops.

## 3. Experimental details:

Lead-free $(Ba_{0.95}Ca_{0.05})(Ti_{0.89}Sn_{0.11})O_3$ ceramic (BCTSn) was obtained via a sol-gel reaction technique. Minutely, a stoichiometric amount of barium acetate $Ba(CH_3COO)_2$ and calcium acetate $Ca(CH3COO)_2$ were dissolved in acetic acid (Solution 1) and tin chloride $(SnCl_2.2H_2O)$ in 2-methoxyethanol(Solution 2) separately. The obtained solutions (1 and 2) were mixed at room temperature. Then, the stoichiometric volume of titanium isopropoxide was added to the mixture solutions. The obtained mixture was stirred for 1 hour, and then ammonia was dropped wisely to the solution to increase the pH value to 7 until the solution became transparent. After the aging step for 15h, the gel was dried at ~80 °C, and the resulting powder was calcined at 1050°C for 4h based on the TGA results (Fig. 2a). The process flow chart is shown in Fig. 1. The resulting calcined powder was uni-axially pressed, without using any binder, to give a pellet sintered at 1350°C for 7h. The pellet was coated with silver-paste electrodes on both surfaces for electrical measurements.

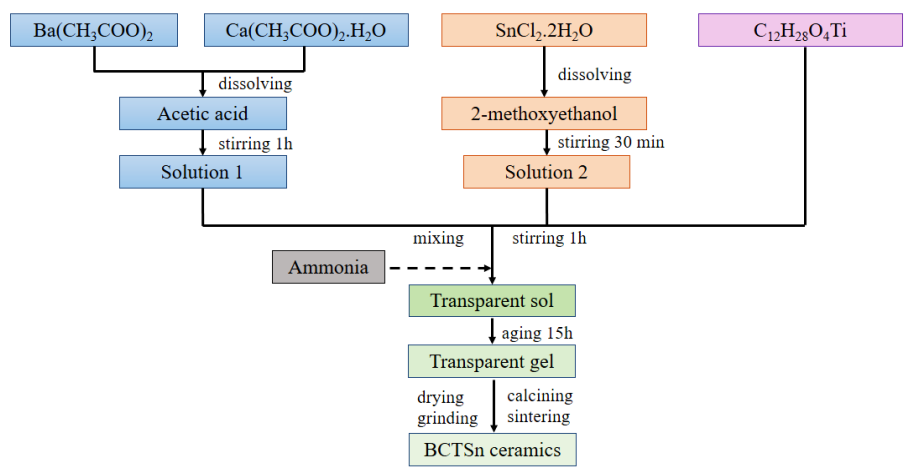

**Fig.1** The synthesis process of BCTSn ceramic by the sol-gel method.

## 4. Results and discussion

### *4.1 Structural and microstructural analysis*

Fig. 2a shows the TGA curve of the uncalcined BCTSn powder. The thermogram represents three main weight losses. From ambient temperature up to around 500 °C, two losses (23% and

16%) correspond to non-structural water and decomposition of thermally unstable organic compounds. The third loss (6%) is from 500 °C to 1000 °C that can be attributed to the undecomposed organic substances and the initial formation of the BCTSn perovskite phase [29]. The loss curve remains stable, which denotes that the crystallization temperature of the desired phase starts from 1000 °C. This calcination temperature is less than those reported using the solid-state method [30]. Fig. 2b shows the XRD pattern at room temperature of BCTSn powder calcined at 1050°C for 4h. All characteristic peaks of the perovskite phase are well observed without any impurity or secondary crystalline phase, suggesting that $Ca^{2+}$ and $Sn^{4+}$ are well diffused into the $BaTiO_3$ lattice forming a solid solution.

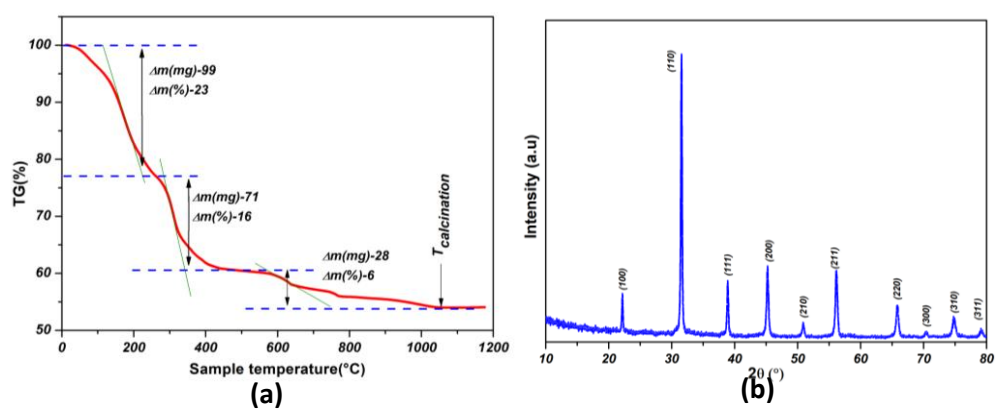

**Fig.2** (a) TGA curve of the BCTSn powder. (b) Room-temperature XRD pattern of powder.

The XRD pattern was refined using the Rietveld refinement program via FullProf software basing on the combination of tetragonal structure with the space group (P4mm) and orthorhombic structure (Amm2). The result proves an excellent agreement between the calculated and the observed diffractograms (Fig. 3a). The deconvoluted peaks between $2\theta \approx 44$–$46°$ and $2\theta \approx 65$–$66°$ indicate the coexistence of tetragonal and orthorhombic phases (inset Fig.3a) as reported by Mezzourh et al. [31]. The crystallite size was determined using Scherrer's formula [31] and found to be 33.207 nm. The crystal structure parameters and the quality-of-fit measure ($\chi^2$) extracted from the Rietveld refinements are shown in Table 1.

**Table 1** Refined structural parameters for BCTSn ceramic at room temperature.

| Structure | Unit cell parameters | | | | | Phase compositions (%) | $\chi^2$ | Average crystalline size(nm) By XRD | Average grain size(µm) by SEM |
| --- | --- | --- | --- | --- | --- | --- | --- | --- | --- |
| | a(Å) | b(Å) | c(Å) | Angle (°) | V(Å³) | | | | |
| **Tetragonal P4mm** | 4.0074 | 4.0074 | 4.0133 | $\alpha = \beta = \gamma = 90$ | 64.449 | 72.92 | 2.703 | 33.207 | 28.13 |
| **Orthorhombic** | 4.0040 | 5.7261 | 5.7261 | | | | | | |

| | | | | | | |
|---|---|---|---|---|---|---|
| *Amm2* | | | | 130.149 | 27.08 | |

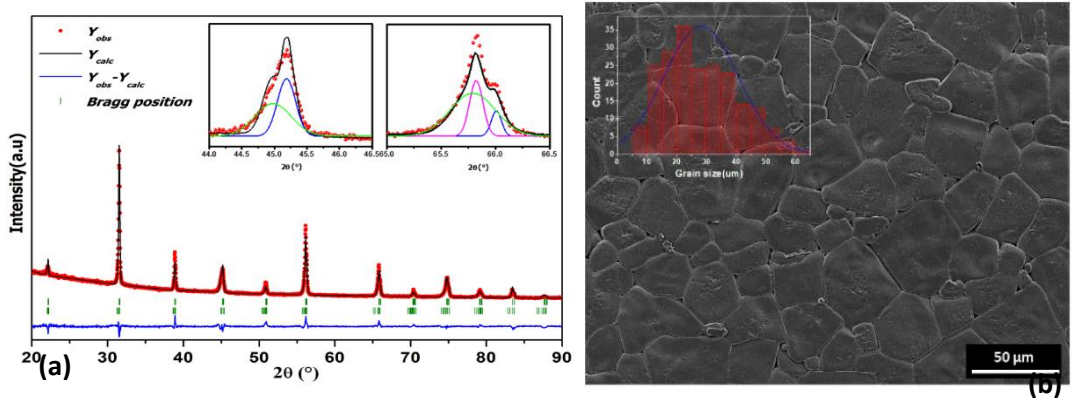

**Fig.3** (a) Rietveld fitted X-ray diffraction patterns of BCTSn ceramic; (b) SEM micrograph of ceramic, the inset to the image shows a grain size distribution.

Fig. 3b shows the SEM micrograph and grain size distribution of BCTSn ceramic sintered at 1350 °C/7h. As it can be seen, large microstructures are observed with well-defined grain boundaries. The grain-size distribution estimated by ImageJ® software is found to be around 28±0.7µm. The apparent density of ceramic is 5.73 g cm$^{-3}$, along with a relative density of 92%. The elemental composition was verified by the EDX analysis. Qualitative analysis reveals the presence of Ba, Ca, Ti, Sn and O elements. The obtained atomic percentage of elements is in good agreement to the expected ratios (C: 13.73%; O: 49.68%; Sn: 2.12%; Ca: 1.03%; Ba: 15.25% and Ti: 18.19%).

*4.2. Dielectric properties*

The temperature dependence of the dielectric constant $\varepsilon_r$ and dielectric loss factor tanδ for BCTSn at the selected frequencies from 100 Hz to 100 kHz is shown in Fig. 4a. The curves at different frequencies display three distinct peaks at 12, 31 and 47°C that correspond to the phase transitions of R-O ($T_{R-O}$), O-T ($T_{O-T}$) and T-C phase ($T_C$), respectively (see the inset to Fig. 4a). This result is in good agreement with DSC measurements that show three exothermic peaks at the same transition temperature values ($T_{R-O}$; $T_{O-T}$, and $T_C$) (Fig. 4b). Compared to BT, the substitution of Ti$^{4+}$ by Sn$^{4+}$ ions shifts $T_{R-O}$ and $T_{O-T}$ to higher temperatures and reduces $T_C$, while the substitution of Ba$_{2+}$ by a small amount of Ca$^{2+}$ does not affect the $T_C$ [32]. It is observed that the values of $\varepsilon_r$ are higher at low frequencies and decrease with the frequency increase. Some of the polarization mechanisms do not contribute when the frequency of the electric field increases, which leads to a decrease in the total polarization, and therefore the decrease in $\varepsilon_r$ [33]. It is worth noting that the frequency increment does not affect the Tc value, confirming that the BCTSn is a classical ferroelectric material without any relaxation behavior[34]. In addition, the maximum dielectric constant at 1 kHz is about 11979, which is much higher than that of $Ba_{0.95}Ca_{0.05}Ti_{0.89}Sn_{0.10}O_3$ ceramic elaborated by a solid-state method[35].

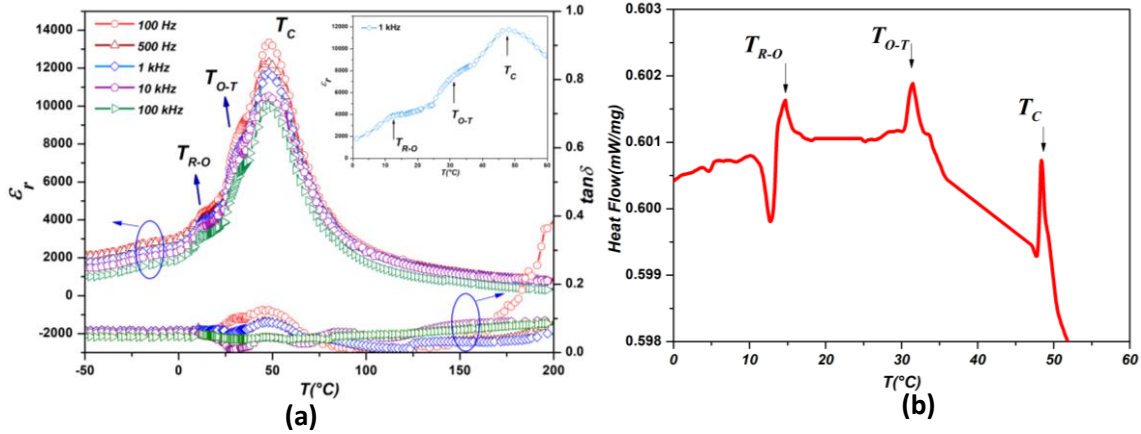

**Fig.4**(a) Temperature-dependence of the dielectric constant and dielectric loss at different frequencies for BCTSn ceramic sintered at 1350/7h (inset: close up of the dielectric constant curve for 1kHz).

(b) Heat flow of BCTSn.

To quantify the critical behavior of BCTSn, the Curie-Weiss law was initially used (Eq. 1) [33],

$$\frac{1}{\varepsilon_r} = \frac{T-T_0}{C} \quad (T > T_0). \quad \textit{(Eq. 1)}$$

Where $T_0$ is the Curie–Weiss temperature and C is the Curie-Weiss constant.

The value of C is about $1.13 \times 10^5$ K (at 1 kHz) which means that BCTSn is displacive-type ferroelectric. The fitting results obtained by equation 1 are shown in table 2. The dielectric constant curve deviates from the Curie-Weiss law at $T_{dev} > T_C$. The degree of this deviation is estimated by equation 2:

$$\Delta T_m = T_{dev} - T_m \quad \text{(Eq. 2)}$$

This deviation is often observed in ferroelectric ceramics and explained by the movement of domain walls [33]. In the BCTSn ceramic, the substitution of $Ti^{4+}$ by $Sn^{4+}$ leads to the distortion of the Ti-O bonds since the ionic radius of $Sn^{4+}$ is bigger than that of $Ti^{4+}$, and therefore $Sn^{4+}$ ions have less free space to move in the oxygen octahedron. The spatial fluctuations of the bonds cause fluctuations in polar density, hence the Curie peak is enlarged [36]. The diffuseness of the critical behavior of BCTSn can be described empirically by the equation 3 [37]:

$$\frac{1}{\varepsilon_r} - \frac{1}{\varepsilon_m} = \frac{(T-T_m)^\gamma}{C} \quad (1 < \gamma < 2) \quad \text{(Eq. 3)}$$

Where $\varepsilon_m$ is the maximum of the dielectric constant, $T_m$ is the temperature that corresponds to $\varepsilon_m$, and γ indicates the character of the phase transition. Thus, γ=1 refers to normal ferroelectrics, and γ=2 corresponds to relaxor ferroelectrics, as described by Hunpratub et al. [38].

For *1 <γ<2*, the diffuse phase transition is incomplete[38]. The value of γ at 1 kHz is extracted from the fitting curve of $ln\left(\frac{1}{\varepsilon_r} - \frac{1}{\varepsilon_m}\right)$ versus $ln(T - T_m)$ and found to be 1.428 that identifies incomplete diffuse phase transition behavior. The results of the dielectric studies are summarized in table 2.

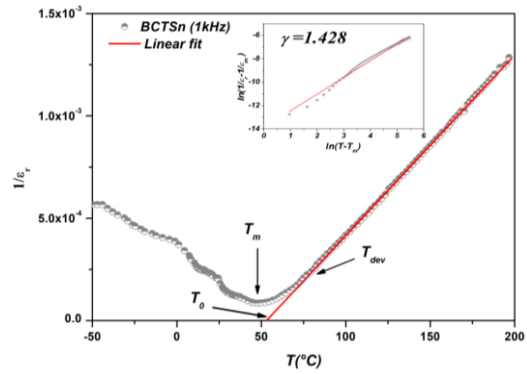

**Fig.5** Curie–Weiss fitting curves of BCTSn sample, the inset is fitted to modified Curie-Weiss law with slope γ=1.428.

**Table 2** Dielectric properties at 1 kHz of the BCTSn ceramic sintered at 1350 °C/7h

|  | $\varepsilon_m$ | tanδ | $T_0$ (°C) | $C \times 10^5$ (K) | $T_m$ (°C) | $T_{dev}$ (°C) | $\Delta T_m$ (°C) | γ |
|---|---|---|---|---|---|---|---|---|
| **BCTSn** | 11979 | 0.105 | 53 | 1.13 | 47 | 79 | 32 | 1.428 |

*4.3 Ferroelectric properties and Energy storage performances:*

The electric field dependent polarization characteristics (P-E loops) are shown in Fig.6a. The polarization loop was measured for various temperatures under the applied electric field of 22.5Kv/cm at 20 Hz. Below $T_C$, moderately saturated hysteresis loops describe the ferroelectric character of BCTSn ceramic. On heating, the hysteresis loops become more slanted and slim that is more desirable for energy storage applications [39]. However, the hysteresis does not disappear above T$_C$, due to the diffuse phase transition behavior in good agreement with dielectric results. Above 383 K, the polarization shows a linear behavior, indicating the appearance of the paraelectric phase. At room temperature, the maximal polarization ($P_{max}$), the remnant polarization ($P_r$), the charge storage density ($Q_C = P_{max} - P_r$) and the coercive field ($E_c$) are found to be 18.93 µC cm$^{-2}$, 6.77 µC cm$^{-2}$, 12.16 µC cm$^{-2}$ and 1.68 kV cm$^{-1}$, respectively, and decrease with increasing the temperature, confirming the ferroelectric-paraelectric phase transition (Fig. 6b). It should be noted that BCTSn ceramic exhibits a low E$_c$ that makes energy barriers lower and facilitate the domain switching during the poling process due to multiphase coexistence[40].

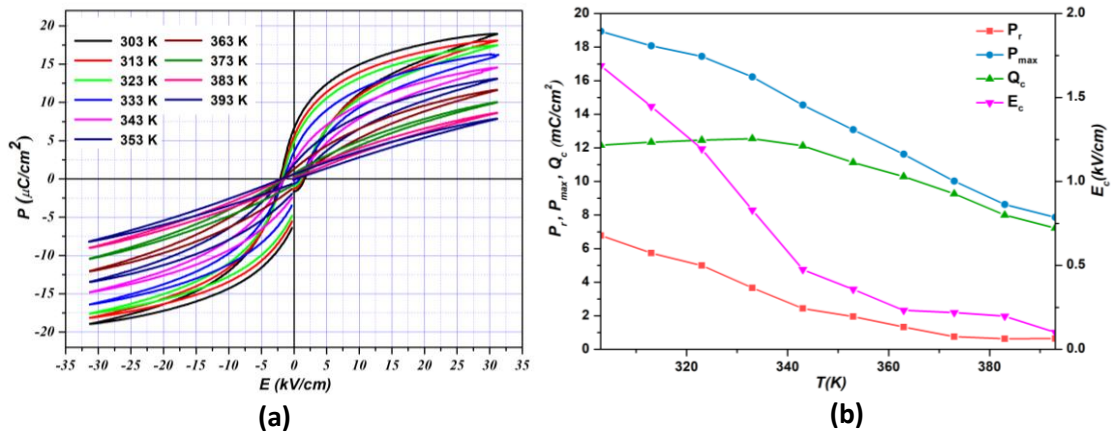

**Fig.6** (a) P-E hysteresis loops of the BCTSn samples. (b) Thermal evolution of P$_r$, P$_{max}$, Q$_c$, and E$_c$.

In order to evaluate the energy storage properties used for energy storage applications, the recoverable energy density W$_{rec}$, total energy density W$_{tot}$ and the energy storage efficiency η are calculated by the equations 4, 5 and 6:

$$W_{tot} = \int_0^{Pmax} E\ dP \quad (Eq.\ 4)$$

$$W_{rec} = \int_{Pr}^{Pmax} E\ dP \quad (Eq.\ 5)$$

$$\eta(\%) = \frac{W_{Rec}}{W_{tot}} \times 100 = \frac{W_{rec}}{W_{rec} + W_{loss}} \times 100 \quad (Eq.\ 6)$$

P$_{max}$, E, P$_r$ and W$_{loss}$ are maximum polarization, applied electric field, remnant polarization and energy loss density, respectively. The W$_{tot}$, W$_{rec}$ and W$_{loss}$ are calculated by measuring the internal and external area of the polarisation versus electric field (P–E) curves as mentioned in Fig. 7a.

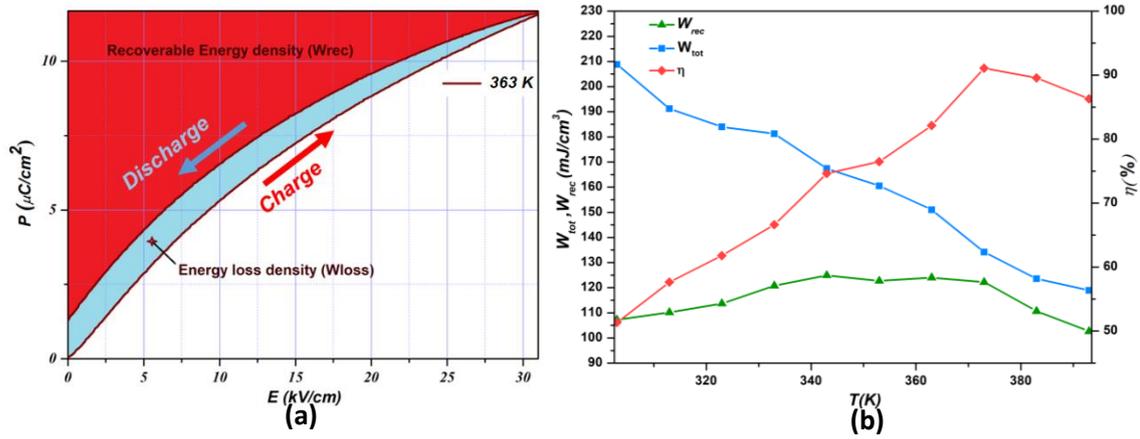

**Fig.7** (a) Schematic illustration of the calculation of $W_{rec}$ and $W_{loss}$ parameters. (b) Energy storage performances in the BCTSn ceramic.

Fig. 7b shows the thermal evolution of $W_{rec}$, $W_{tot}$ and $\eta$ that found to be 107.25, 208.83 mJ cm$^3$ and 51% at room temperature, respectively. However, with temperature increasing, $W_{tot}$ decreases and attains 118.89 mJ cm$^{-3}$ at 393K. $W_{rec}$ slowly raises and remains stable at 124 mJ cm$^{-3}$ in the temperature range of 343-373K and then decreases. Accordingly, $\eta$ upsurges to reach 91.07 % at 373 K. Therefore, it is a promising material for energy storage applications. It is worth noting that no research on the energy storage properties of BCTSn has been investigated.

Table 3 presents the energy storage properties of some lead-free ceramics materials reported in the literature. Using an electric field of 20 kV cm$^{-1}$, Wei Ca reported a recovered energy density of 38.6 mJ cm$^{-3}$ and storage efficiency of 33.9% in the $Ba_{0.85}Ca_{0.15}Zr_{0.1}Ti_{0.9}O_3$ ceramic (BCZT) with almost the same grain size (32μm) [41]. Meanwhile, by the same synthesis method we used, Mezzourh et al. found a recovered energy density of 10.4 mJ cm$^{-3}$ and storage efficiency of 59.9 % under 8.75 kV cm$^{-1}$ in $Ba_{0.85}Ca_{0.15}Zr_{0.1}Ti_{0.9}O_3$ ceramic [31]. Furthermore, our previous work showed that BCZT ceramic achieved at 300K an energy storage density of 367.2 mJ cm$^{-3}$ at 60 kV cm$^{-1}$ [22]. It should be noted that our BCTSn ceramic exhibited higher energy storage performances than those found in the BCZT ceramics synthesized via various methods.

**Table 3** Various BT based ceramics along with their respective energy storage properties.

| Ceramic | method | $W_{tot}$ (mJ/cm$^3$) | $W_{rec}$ (mJ/cm$^3$) | $\eta$ (%) | E (kV/cm) | T (K) | Refs |
|---|---|---|---|---|---|---|---|
| $Ba_{0.95}Ca_{0.05}Ti_{0.89}Sn_{0.11}O_3$ | Sol gel | 208.83 | 107.25 | 51.36 | 31 | 303 | This work |
| $Ba_{0.95}Ca_{0.05}Ti_{0.89}Sn_{0.11}O_3$ | Sol gel | 134.18 | 122.20 | 91.07 | 31 | 373 | This work |
| $Ba_{0.95}Ca_{0.05}Ti_{0.89}Sn_{0.11}O_3$ | Sol gel | 167.37 | 124.91 | 74.63 | 31 | 343 | This work |
| BaTi$_{0.89}$Sn$_{0.11}$O$_3$ | Solid state | 67.9 | 65.1 | 95.87 | 25 | 373 | [42] |
| Ba$_{0.95}$Ca$_{0.05}$Zr$_{0.20}$Ti$_{0.80}$O$_3$ | Solid state | 569.4 | 410 | 72 | 120 | 303 | [43] |
| Ba$_{0.85}$Ca$_{0.15}$Zr$_{0.10}$Ti$_{0.90}$O$_3$ | Solid state | 113.8 | 38.6 | 33.9 | 20 | 298 | [41] |
| Ba$_{0.85}$Ca$_{0.15}$Zr$_{0.10}$Ti$_{0.90}$O$_3$ | Sol gel | 17.36 | 10.40 | 59.9 | 8.75 | 303 | [31] |
| Ba$_{0.85}$Ca$_{0.15}$Zr$_{0.10}$Ti$_{0.90}$O$_3$ | Hydrothermal | 546.1 | 367.2 | 67.2 | 60 | 300 | [22] |
| Ba$_{0.7}$Ca$_{0.3}$TiO$_3$ | Solid state | - | | 58 | 50 | 303 | [44] |

*4.5 Indirect electrocaloric measurements:*

The indirect experimental method evaluated the EC effect by exploiting P(E,T) results. The reversible adiabatic temperature change ΔT was estimated from the Maxwell equation 7. A fifth-order polynomial fitting of the upper polarization branches was carried out at each fixed applied electric field.

$$\Delta T = - \int_{E2}^{E1} \frac{T}{\rho C p} \left(\frac{\partial P}{\partial T}\right)_E dT \quad (Eq.\ 7)$$

Here $\rho$ is the density of ceramic and Cp is the specific heat of the materials estimated at 0.39J g$^{-1}$K$^{-1}$[45].

Fig. 8a and b depicts the thermal evolution of the temperature change (ΔT) and the corresponding entropy change (ΔS) at various electric fields of BCTSn ceramic. Two large peaks corresponding to the O-T and T-C phase transitions are observed at 315K and 337K, respectively, for both ΔT and ΔS and found to be shifted to high temperature comparing with dielectric results. Such difference could result from different thermometry used in each technique. Same behavior was observed in x(Ba$_{0.7}$Ca$_{0.3}$)TiO$_3$ (1-x) Ba(Sn$_{0.11}$Ti$_{0.89}$)O$_3$ with (x=0,0.1,0.2 and 0.3) reported by Chunlin Zhao et al [46]. Our BCTSn sample presents a high ECE response and ΔS of 0.804 K and 0.87 J kg$^{-1}$ K$^{-1}$ at 30 kVcm$^{-1}$, respectively, around the ferroelectric-paraelectric transition. The thermal evolution of electrocaloric responsivity ζ = ΔT/ΔE under different applied electric fields is shown in Figures 8c and d. ζ follows the same evolution as that of ΔT and ΔS and achieved a high value of 0.268 K mm kV$^{-1}$ at 30 kV cm$^{-1}$. The obtained results (ΔT = 0.54 K and ζ = 0.275 K mm kV$^{-1}$ at 335 K under 20kV cm$^{-1}$) are similar to those found in the Ba$_{0.94}$Ca$_{0.06}$Ti$_{0.89}$Sn$_{0.11}$O$_3$ composition elaborated via the solid-state method (ΔT = 0.55 K and ζ = 0.28 K mm kV$^{-1}$ at 320 K under 20kV cm$^{-1}$) by Wang et al. [19], and more comparisons with other BT based ceramics are presented in table 4. The outstanding

EC response takes place in the BCTSn sample mainly due to the coexistence of multiple phases at this composition. In reality, the contribution of Ferro-Ferro transitions results in significant intrinsic ferroelectric polarization, which can improve entropy changes via the disordered/ordered polar dipoles. Thus, a high ΔT comparing with other doped BT using single Ferro-Para transition could be achieved [46]–[48]. Therefore, designing multiphase coexistence in EC material systems is a good way to improve their EC properties [45], [49], [50]. In addition, various factors contribute to the EC responses, such as grain size, synthesis conditions and an applied electric field.

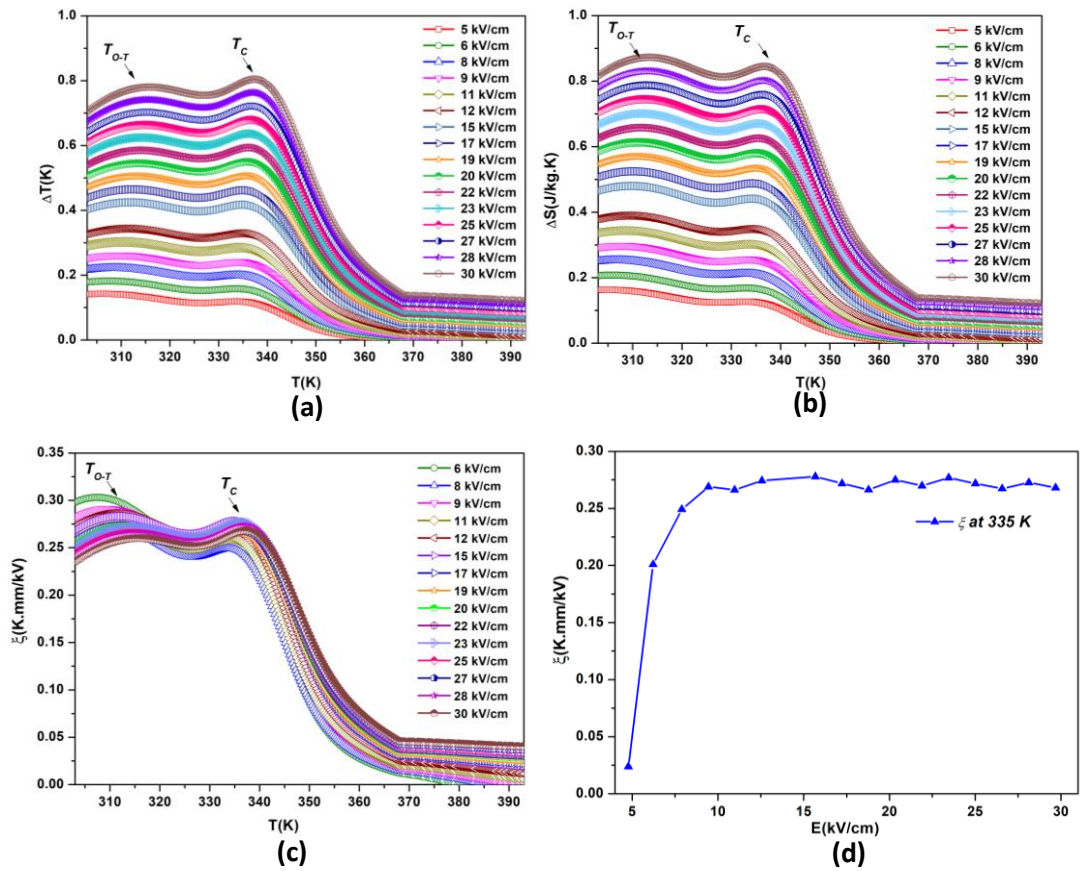

**Fig. 8** Temperature-dependence of (a) Δ$T$, (b) Δ$S$ and (c) ξ at different applied electric fields in the BCTSn ceramic. (d) Electric field-dependence of ξ at 335 K.

**Table 4** Comparison of the electrocaloric properties of the BCTSn ceramics with other BT based ceramics reported in the literature using the indirect method

| Ceramic | $T(K)$ | $ΔT(K)$ | $ΔE$ (kV cm$^{-1}$) | ξ(K mm kV$^{-1}$) | Ref |
|---|---|---|---|---|---|
| Ba$_{0.95}$Ca$_{0.05}$Ti$_{0.89}$Sn$_{0.11}$O$_3$ | 335 | 0.807 | 30 | 0.268 | This work |
| Ba$_{0.95}$Ca$_{0.05}$Ti$_{0.89}$Sn$_{0.11}$O$_3$ | 335 | 0.54 | 20 | 0.275 | This work |
| Ba$_{0.95}$Ca$_{0.05}$Ti$_{0.89}$Sn$_{0.11}$O$_3$ | 335 | 0.41 | 15 | 0.278 | This work |
| Ba$_{0.94}$Ca$_{0.06}$Ti$_{0.90}$Sn$_{0.10}$O$_3$ | 320 | 0.55 | 20 | 0.28 | [19] |
| Ba$_{0.94}$Ca$_{0.06}$Ti$_{0.95}$Sn$_{0.05}$O$_3$ | 358 | 0.59 | 20 | 0.30 | [19] |
| BaTi$_{0.89}$Sn$_{0.11}$O$_3$ | 373 | 0.57 | 25 | 0.23 | [42] |
| 0.7BaZr$_{0.2}$Ti$_{0.8}$O$_3$-0.3Ba$_{0.7}$Ca$_{0.3}$TiO$_3$ | 328 | 0.3 | 20 | 0.15 | [51] |
| Ba$_{0.95}$Ca$_{0.10}$Ti$_{0.95}$Sn$_{0.05}$O$_3$ | 398 | 0.34 | 12.69 | 0.27 | [52] |
| Ba$_{0.80}$Ca$_{0.20}$Zr$_{0.15}$Ti$_{0.85}$O$_3$ | 307 | 0.235 | 22.3 | 0.105 | [53] |
| Ba$_{0.30}$Ca$_{0.10}$Zr$_{0.05}$Ti$_{0.95}$O$_3$ | 392 | 0.565 | 30 | 0.188 | [53] |
| Ba$_{0.92}$Ca$_{0.08}$Zr$_{0.05}$Ti$_{0.92}$O$_3$ | 320 | 0.38 | 15 | 0.25 | [53] |
| Ba$_{0.65}$Sr$_{0.35}$TiO$_3$ | 289 | 0.23 | 10 | 0.23 | [54] |
| Ba$_{0.65}$Sr$_{0.35}$TiO$_3$ | 296 | 0.42 | 20 | 0.21 | [54] |
| Ba$_{0.65}$Sr$_{0.35}$TiO$_3$ | 303 | 2.1 | 90 | 0.23 | [55] |

For industrial applications, the efficiency of ECE of BCTSn sample was examined via the coefficient of performance COP as defined by equation 8 [56],

$$COP = \frac{|Q|}{|W_{tot}|} = \frac{|T\Delta S|}{|W_{tot}|}. \quad (Eq.\ 8)$$

Where Q represents the isothermal heat.

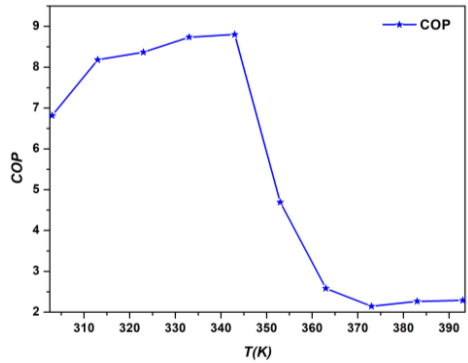

**Fig. 9** Temperature-dependence of the COP.

The COP calculated from 303 to 393K under 30kV cm$^{-1}$ is shown in Fig. 9. It increases slightly with temperature and reaches a maximum value of 8.8 at 342K and then decreases above the transition temperature. Our COP value is larger than those found in (1−x)K$_{0.5}$Na$_{0.5}$NbO$_3$-xLaNbO$_3$ ceramic (COP=4.16) under 50kVcm$^{-1}$[56], and in Ba$_{0.85}$Ca$_{0.15}$Zr$_{0.10}$Ti$_{0.90}$O$_3$ ceramic upon 55 kV cm$^{-1}$ (COP=6.29 at 365K)[57].

## 5 Conclusion

In summary, the environment-friendly Ba$_{0.95}$Ca$_{0.05}$Ti$_{0.89}$Sn$_{0.11}$O$_3$ ceramic was obtained by the sol-gel method. Structure, dielectric, phase transitions, energy storage and electrocaloric properties were investigated. As a result, BCTSn showed enhanced energy storage performances of W$_{rec}$=124 mJ cm$^{-3}$ and η = 91.07% at 373 K under an electric field of 30 kV

cm$^{-1}$. Moreover, significant electrocaloric response was achieved (ΔT = 0.807 K, ΔS = 0.844 J kg$^{-1}$ K$^{-1}$, ζ = 0.268 K mm kV$^{-1}$ under 30 kV cm$^{-1}$), with an encouraging COP value of 8.8 at 343 K. Thus, lead-free BCTSn has a significant potential to be considered as a good candidate for environmentally friendly refrigeration and high energy storage applications.